\documentclass{PoS}

\usepackage{amsmath,amssymb,url}

\title{Theory of radiative $B$ decays}

\ShortTitle{Theory of radiative $B$ decays}

\author{\speaker{Gil Paz}\\
Department of Physics and Astronomy \\
Wayne State University\\
Detroit, Michigan 48201, USA \\
                      E-mail: \email{gilpaz@wayne.edu}}

\abstract{This talk discusses the theory of inclusive and exclusive radiative B decays, emphasizing the interplay of perturbative and non-perturbative effects and the importance CP and isospin asymmetries.}

\FullConference{ 9th International Workshop on the CKM Unitarity Triangle\\
                 28  November - 3 December 2016\\
                 Tata Institute for Fundamental Research (TIFR), Mumbai, India}

\begin{document}

\section{The Big Picture}
Why radiative $B$ decays?
\begin{itemize}
\item Radiative $B$ decays are... an important probe of New Physics. The process $b\to s \gamma\,$ is a flavor changing neutral current (FCNC). In the Standard Model (SM) there are no FCNCs at tree level. They only arise at a loop level. As a result, $b\to s \gamma$ can have contributions from new physics, e.g. Supersymmetry. Thus radiative $B$ decays constrain many models of new physics.

\item Radiative B decays are... theoretically clean. Since the $b$-quark mass ($m_b\sim5$ GeV) is much larger than $\Lambda_{\mbox{\scriptsize  QCD}}\sim 0.5$ GeV, observables can be expanded as a power series in $\Lambda_{\mbox{\scriptsize  QCD}}/{m_b}\sim 0.1$. For example, the $Q_{7\gamma}-Q_{7\gamma}$ contribution to $\Gamma(\bar B\to X_q \gamma)$ is known to ${\cal O}\left({ \Lambda^5_{\mbox{\scriptsize  QCD}}}/{m^5_b}\right)$ \cite{Mannel:2010wj}. This allows to systematically control non-perturbative effects.  

\item Radiative B decays are... theoretically interesting. They test basic quantum field theory tools such as the operator product expansion (OPE) and factorization theorems that are important in other areas of physics. They also open a window to non-perturbative physics. For example, the photon spectrum of $\bar B\to X_s \gamma$ decays measured by CLEO \cite{Chen:2001fja}, BaBar \cite{Lees:2012ufa}, and Belle \cite{Belle:2016ufb} collaborations is (at leading twist) the $B$-meson parton distribution function.   

\item Finally, radiative B decays have... a large impact. Of the CLEO collaboration top cited papers, papers on radiative $B$ decays are at the first ($\bar B\to X_s \gamma$ \cite{Alam:1994aw}), third ($\bar B\to K^* \gamma$ \cite{Ammar:1993sh}) and fourth ($\bar B\to X_s \gamma$ \cite{Chen:2001fja}) place\footnote{A paper concerning the CLEO-II detector \cite{Kubota:1991ww} is at the second place.} . Similarly a paper on $\bar B\to X_s \gamma$ \cite{Abe:2001hk} is the fourth most cited of the Belle collaboration. For the BaBar collaboration the most highly cited paper on radiative $B$ decays ($\bar B\to X_s  \ell^+ \ell^-$ \cite{Aubert:2004it}) is ``only" at the 21st place\footnote{All citation information is based on inspirehep \cite{inspirehep} citation count at the time of CKM 2016.}. Furthermore, theory papers on radiative $B$ decays have hundreds of citations. 
\end{itemize}
Radiative $B$ decays exhibit the interplay of perturbative (short distance) and non-perturbative (long distance) physics, where the latter can be quite complex in structure. The perturbative physics is described by the effective Hamiltonian  \cite{Buras:1998raa} 
\begin{equation}
{\cal H}_{\mbox{\scriptsize eff}}=\frac{G_F}{\sqrt{2}}\sum_{p=u,c}\lambda_p\left( C_1 Q_1^p+C_2 Q_2^p+\sum_{i=3,...,10}C_iQ_i+C_{7\gamma}Q_{7\gamma}+C_{8g}Q_{8g}\right).
\end{equation}
The most important operators for radiative $B$ decays are 
\begin{eqnarray}
Q_{7\gamma}&=&\frac{-e}{8\pi^2}m_b\bar s
\sigma_{\mu\nu}(1+\gamma_5)F^{\mu\nu}b,\nonumber\\
Q_{8g}&=&\frac{-g_s}{8\pi^2}m_b\bar s
\sigma_{\mu\nu}(1+\gamma_5)G^{\mu\nu}b,\nonumber\\
Q_1^q&=&(\bar q b)_{V-A}(\bar s q)_{V-A}\quad (q=u,c).
\end{eqnarray}
The Wilson coefficients $C_i$ known at next-to-next-to-leading order (NNLO). Non-perturbative effects arise at ${\cal O}\left(\Lambda_{\mbox{\scriptsize  QCD}}/m_b\right)$.

In the following we survey the current theoretical status of inclusive and exclusive radiative $B$ decays. Among the themes common to both are the interplay of perturbative (short distance) and non-perturbative (long distance) physics and the importance of asymmetries: isospin asymmetries ($\bar B\to X_s \gamma, \bar B\to K^*\gamma, \bar B\to \rho\gamma$), CP asymmetries ($\bar B\to X_s \gamma, \bar B\to K^*\gamma, \bar B\to \rho\gamma$) and even isospin difference of CP asymmetries \cite{Benzke:2010tq,Lees:2014uoa}. While not discussed here, one should mention radiative decays that contain more than one photon: $B\to \gamma\gamma$ and $B\to X_q\gamma\gamma$. These were not observed yet.

\section{Inclusive $\bar B\to X_q \gamma$ }
The most recent theoretical prediction for inclusive $\bar B\to X_q \gamma$ decays are $\mbox{Br}(\bar B\to X_s \gamma)=(3.36 \pm 0.23) \times 10^{-4}$ and $\mbox{Br}(\bar B\to X_d \gamma)=(1.73^{+0.12}_{-0.22}) \times 10^{-5}$ \cite{Misiak:2015xwa}. These should be compared to the experimental values of $\mbox{Br}(\bar B\to X_s \gamma)=(3.43 \pm 0.21\pm0.07) \times 10^{-4}$ \cite{Amhis:2014hma} and 
$\mbox{Br}(\bar B\to X_d \gamma)=(1.41\pm 0.57) \times 10^{-5}$ \cite{delAmoSanchez:2010ae,Misiak:2015xwa}. It should be noted that these measurements are obtained by extrapolating the experimental measurement with a photon energy cut in the range of $\left[1.7,2.0\right]$ GeV to 1.6 GeV in order to compare it to theory. 

At leading power in $\Lambda_{\mbox{\scriptsize  QCD}}/m_b$ we have $\Gamma(\bar B\to X_s \gamma)=\Gamma(b\to s \gamma)$, namely the decay is equal to that of a free $b$-quark. Since $\Gamma\propto |{\cal H}_{\mbox{\scriptsize eff}}|^2$, pairs of operators contribute to the decay rate. The contribution of  $Q_{7\gamma}-Q_{7\gamma}$ and $Q_{7\gamma}-Q_{8g}$ is  known at NNLO, while $Q_{1,2}-Q_{7\gamma}$ are known at NNLO for two $m_c$ limits. Interpolation in $m_c$ leads to $\pm3\%$ perturbative uncertainty. Future improvement requires calculation at physical $m_c$ which is challenging.

Turning to non-perturbative effects, the $Q_{7\gamma}-Q_{7\gamma}$ contribution obeys a local OPE of the form
\begin{equation}
\Gamma_{77}=\sum_{n=0}^\infty\dfrac1{m_b^n}\sum_k c_{k,n}\langle O_{k,n}\rangle, 
\end{equation} 
where $c_{k,n}$ are perturbative Wilson coefficients and $\langle O_{k,n}\rangle$ are non-perturbative matrix elements of HQET operators. The $n=0$ term gives the free quark contribution discussed above, while the $n=1$ term vanishes. The coefficients  $c_{k,2}$ are known at ${\cal O}(\alpha_s)$ \cite{Ewerth:2009yr}. The coefficients $c_{k,\{3,4,5\}}$ are known at ${\cal O}(\alpha^0_s)$ \cite{Mannel:2010wj} but $\langle O_{k,\{3,4,5\}}\rangle$ are not well known, see \cite{Gambino:2016jkc}. Still, this is arguably the best known OPE prediction. Very recently, even higher dimensional matrix elements were presented in \cite{Gunawardana:2017zix}, see the talk ``Higher dimensional HQET parameters" in these proceedings. Their contribution to $\Gamma_{77}$ is not known yet. 

Non-perturbative effects from other operators are more complicated. These operators lead to resolved photon contributions, namely processes in which the photon is not directly produced. Thus the operator $Q_{8g}$ can contribute to the process  $b\to sg\to s\bar q q\gamma$ and $Q_{1}$ to $b\to s\bar c c\to sg\gamma$. The final state contain more partons and might seem to be highly power suppressed, but a more careful analysis shows that these effects give a  ${\cal O}\left(\Lambda_{\mbox{\scriptsize  QCD}}/m_b\right)$ contribution to the rate. For more details see \cite{Benzke:2010js} and the CKM 2010 talk \cite{Paz:2010wu}. The rate can be written symbolically as $\Gamma\sim \bar{\!\!J} \otimes h$, where  $\,\,\bar{\!\!J}$ are perturbative functions and  $h$ are non-perturbative. These non-perturbative functions must be modeled or extracted from (future) data. At ${\cal O}\left(\Lambda_{\mbox{\scriptsize  QCD}}/m_b\right)$ one finds contributions from $Q_{1}-Q_{7\gamma}$, $Q_{7\gamma}-Q_{8g}$, and $Q_{8g}-Q_{8g}$, leading to a total uncertainty of 5\% on $\Gamma(\bar B\to X_s \gamma)$. This is also the largest uncertainty on the rate \cite{Misiak:2015xwa}.

Can the non-perturbative uncertainty be reduced? For the $Q_{1}-Q_{7\gamma}$ contribution, the improved knowledge of  $\langle O_{k,n}\rangle$ \cite{Gambino:2016jkc} from semileptonic data can better constrain it \cite{in_progress}. A better measurement of the isospin asymmetry between $B^+$ and $\bar B^0$ can better constrain the $Q_{7\gamma}-Q_{8g}$ contribution. Thus data driven methods can better constrain the theoretical uncertainty.  

Resolved photon contributions affect the CP asymmetry in $\bar B\to X_s \gamma$. It changes it from $\sim 0.5\%$ from perturbative effects alone to $\left[-0.6\%,2.8\%\right]$. For details see \cite{Benzke:2010tq} and the CKM 2012 talk \cite{Paz:2012sa}. They also imply a test of new physics.  The isospin difference of CP asymmetries, $\Delta A_{CP}$, is proportional to $\mbox{Im } (C_{8g}/C_{7\gamma})$, which can be non-zero in models of new physics. $\Delta A_{CP}$ was so far measured only by BaBar in 2014: $\Delta A_{CP} = +(5.0 \pm 3.9 \pm1.5)\%$ \cite{Lees:2014uoa}. While consistent with zero, this measurement can be used to constrain  $\mbox{Im } (C_{8g}/C_{7\gamma})$. In particular they find $
-1.64\leq \mbox{Im } (C_{8g}/C_{7\gamma})\leq 6.52$ at  90\% confidence level \cite{Lees:2014uoa}.

Finally, resolved photon effects for the photon spectrum are not known numerically. They are relevant for extraction of HQET parameters and $|V_{cb}|$ and $|V_{ub}|$. Also, the comparison between theory and experiment relays on extrapolation from measured $E_\gamma\sim 1.9$ GeV to $E_\gamma>1.6$ GeV. The issue of extrapolation should be revisited. Both issues can benefit from detailed $E_\gamma$ cut effects. For example, instead of optimizing the $E_\gamma$ cut, it would be useful to have measurements with different values of the photon energy cut. This will allow to test the extrapolation and the resolved photon contributions to the photon spectrum against the data. 

\section{Exclusive $\bar b\to q \gamma$}
The theoretical prediction for decays such as  $b\to s\gamma:$ $B_{(q,s)}\to (K^*,\phi) \gamma\,$  and $b\to d\gamma:$ $B_{(q,s)}\to (\rho/\omega, \bar K^*) \gamma$ also show an interplay of short distance (SD) effects from $Q_{7\gamma}$  (while still requiring $B\to V$ form factors) and  long distance (LD) effects from other operators. In order to reduce hadronic uncertainties one can look at ratios and asymmetries that can be more sensitive to new physics effects.  

As an example for ratios consider $R^{\mbox{\scriptsize}}_{K^* \gamma/\phi \gamma}=\mbox{Br}(B\to K^* \gamma)/\mbox{Br}(B_s\to \phi \gamma)$. The SM prediction from \cite{Lyon:2013gba}  is $R^{\mbox{\scriptsize SM}}_{K^* \gamma/\phi \gamma}=0.78\pm0.18$. This should be compared toe the LHCb measurement of $R^{\mbox{\scriptsize exp}}_{K^* \gamma/\phi \gamma}=1.23\pm0.12$ \cite{LHCb:2012ab,Aaij:2012ita}. 

Other observables that are sensitive to new physics are isospin asymmetries. The SM prediction for the isospin asymmetry in $B$ meson decays to $K^* \gamma$ and $\rho \gamma$ are $\bar{a}_I^{\mbox{\scriptsize SM}}(K^* \gamma)=(4.9\pm2.6)\%$ and 
$\bar{a}_I^{\mbox{\scriptsize SM}}(\rho \gamma)=(5.2\pm2.8)\%$. The former agrees with the experimental value $\bar{a}_I^{\mbox{\scriptsize exp}}(K^* \gamma)=(5.2\pm2.6)\%$ \cite{Amhis:2012bh}, while the latter is in tension with  $\bar{a}_I^{\mbox{\scriptsize exp}}(\rho \gamma)=(30^{+16}_{-13})\%$ \cite{Amhis:2012bh}. 

Belle II can study the time dependent CP asymmetry in $B$ meson decays to $f\gamma$, where $f$ is a CP eigenstate \cite{Atwood:1997zr}. The SM prediction changes from $S^{\mbox{\scriptsize SM, SD}}_{K^*(K_s\pi^0) \gamma}=-2\frac{m_s}{m_b}\sin2\phi_1$ and 
$S^{\mbox{\scriptsize SM, SD}}_{\rho^0(\pi^+\pi^-) \gamma}=0$ from short distance effects alone, to $S^{\mbox{\scriptsize SM}}_{K^*(K_s\pi^0) \gamma}=-(2.3\pm1.6)\%$ and $S^{\mbox{\scriptsize SM}}_{\rho^0(\pi^+\pi^-) \gamma}=(0.2\pm1.6)\%$ when including also long distance effects. This is reminiscent of the inclusive CP asymmetry. 

In principle the photon can have two different polarizations in $B\to V\gamma$ decays. The photon polarization is sensitive to ``right handed" new physics.  LHCb observed up-down asymmetry proportional to the photon polarization in $B^\pm\to K^\pm\pi^\mp\pi^\pm\gamma\,$ \cite{Aaij:2014wgo}. But unfortunately as they say at the end of the paper  ``.. the values for the up-down asymmetry, may be used, if theoretical predictions become available, to determine for the first time a value for the photon polarization, and thus constrain the effects of physics beyond the SM in the $b\to s\gamma$ sector". See also Kou's talk at CKM 2014 \cite{Kou}. Hopefully future measurements of other decay channels could be interpreted in terms of the photon polarization.

\section{Conclusions}
Radiative $B$ decays have a rich structure and show intricate interplay of perturbative and non-perturbative physics. At the same time they are an important probe of new physics. While the theory of radiative $B$ decays is mature, there is still room for improvements. New results from LHCb, and soon Belle II, will motivate further theoretical work.

 \section*{Acknowledgments} 
 I would like to thank the organizers for the invitation to give this talk. The author recently contributed to the Belle II Theory interface Platform (B2TiP) report chapter on Radiative and Electroweak Penguin B Decays (Authors: T. Feldmann, U. Haisch, M. Misiak, G. Paz, E. Kou, R. Zwicky) which has influenced these proceedings \cite{B2TiP}. This work was supported by DOE grant DE-SC0007983.

\end{document}